\pdfoutput=1
\documentclass[aip,preprint,superscriptaddress,onecolumn,showpacs,amsmath,floatfix,citeautoscript]{revtex4-1}
\usepackage{graphicx}
\usepackage{float}
\usepackage{dcolumn}
\usepackage{color}
\usepackage{latexsym,bm}
\usepackage[normalem]{ulem}
\usepackage{multirow}
\usepackage{appendix}
\usepackage{amsmath}
\usepackage{amsfonts}
\usepackage{booktabs}

\begin{document}

\hyphenpenalty=5000
\tolerance=1000

%\title{Equation of States of Beryllium Studied by Deep Learning Molecular Dynamics}
\title{Thermal Transport by Electrons and Ions in Warm Dense Aluminum: A Combined Density Functional Theory and Deep Potential Study}

\author{Qianrui Liu}
\affiliation{CAPT, HEDPS, College of Engineering, Peking University, Beijing 100871, P. R. China}
\author{Junyi Li}
\affiliation{School of Computer Science and Technology, Harbin Institute of Technology (Shenzhen), Shenzhen, Guangdong 518055, China}
  \author{Mohan Chen}
  \email{mohanchen@pku.edu.cn (Corresponding author)}
\affiliation{CAPT, HEDPS, College of Engineering, Peking University, Beijing 100871, P. R. China}
\date{\today}

%%%%%%%%%%%%%%%%%%%%%%%%%%%%%%%%%%%%%%%%%%%%%%%%%%%%%%%%%%%%%%%%%%%%%%%%%%%%%%%%%%%%%%%%%%%%%%%%%%%%%%%%%%%%%%%%%%%%%%%%%%%%%%%%%%%
%%%%%     Title
%%%%%%%%%%%%%%%%%%%%%%%%%%%%%%%%%%%%%%%%%%%%%%%%%%%%%%%%%%%%%%%%%%%%%%%%%%%%%%%%%%%%%%%%%%%%%%%%%%%%%%%%%%%%%%%%%%%%%%%%%%%%%%%%%%%

\begin{abstract}
{
%Quantum-mechanics-based methods such as density functional theory (DFT)
%own the predictive power and are suitable for studying warm dense matter.
%
We propose an efficient scheme, which combines density functional theory (DFT) with deep potentials (DP),
to systematically study the convergence issues of the computed electronic
thermal conductivity of warm dense Al (2.7 g/cm$^3$, temperatures ranging from 0.5 to 5.0 eV)
with respect to the number of $k$-points,
the number of atoms, the broadening parameter,
the exchange-correlation functionals and the pseudopotentials.
Furthermore, the ionic thermal conductivity is obtained by the Green-Kubo method
in conjunction with DP molecular dynamics simulations,
and we study the size effects in affecting the ionic thermal conductivity.
This work demonstrates that the proposed method is efficient in evaluating
both electronic and ionic thermal conductivities of materials.
}

\end{abstract}
%\pacs{}
\maketitle

\section{Introduction}
Warm dense matter (WDM) is a state of matter
lying between condensed matter and plasma,
which consists of strongly coupled ions and partially degenerated electrons.
WDM exists in the interior of giant planets~\cite{99S-Guillot,12AJ-Nettelmann} or the crust of white dwarf and neutron stars~\cite{80AJSS-Wesemael,09AJ-Daligault}
and can be generated through laboratory experiments such as diamond anvil cell~\cite{96N-Loubeyre},
gas gun~\cite{96L-Weir,06RPP-Nellis} and high power lasers.~\cite{97PP-Cauble,06RPP-Nellis}
WDM also plays an important role in the inertial confinement fusion (ICF).~\cite{95PP-Lindl}
In this regard, it is crucial to understand properties of WDM such as
the equation of state, optical and transport properties.
Due to the lack of experimental data, quantum-mechanics-based simulation methods
such as Kohn-Sham density functional theory (KSDFT),~\cite{64PR-Hohenberg,65PR-Kohn}
orbital-free density functional theory (OFDFT),~\cite{02TMCPC-Wang}
and path-integral monte carlo~\cite{84B-Pollock,86L-Ceperley,13L-Brown,15L-Militzer}
have emerged as ideal tools to study WDM.~\cite{08B-Holst,05B-Recoules,13PP-Wang,15L-Militzer,20PP-Bonitz}
Thermal conductivity is one of the upmost important properties of warm dense matter.
For example, it is widely studied in the modeling the interior structure of planets.~\cite{68AJ-Hubbard,74I-Seigfried,03PEPI-Labrosse}
It is a key parameter in the hydrodynamic instability growth
on the National Ignition Facility capsule design.~\cite{98PP-Marinak}
It also plays a key role in simulations of the interactions between laser and a metal target.~\cite{03B-Ivanov}
Besides, the thermal conductivity largely affects predictions of ICF implosions in hydrosimulations.~\cite{14E-Hu}

The thermal conductivity includes both electronic and ionic contributions.
The Kubo-Greenwood (KG) formula~\cite{57JPSJ-Kubo,58PPS-Greenwood}
has been widely applied to study the electronic thermal conductivity $\kappa_{e}$
of liquid metals and WDM.~\cite{02E-Desjarlais,13CMS-Knyazev,14PP-Knyazev,14E-Hu,12AJSS-Martin,05B-Recoules,11PP-Flavien,12B-Vlcek,12PP-Starrett}
Typically, the $\kappa_{e}$ value is averaged over results from several atomic configurations,
which are selected from first-principles molecular dynamics (FPMD) simulations.
However, it is computationally expensive to perform FPMD simulations for large systems,
especially for WDM when the temperatures are high.~\cite{10L-Hu,11PP-Flavien,13PP-Wang,14E-Sheppard}
Besides, the size effects may substantially affect $\kappa_{e}$,
as has been demonstrated in previous works.~\cite{11B-Pozzo,11PP-Flavien,13CMS-Knyazev}
For computations of the ionic thermal conductivity $\kappa_I$, we focus on utilizing the
Green-Kubo (GK) formula,~\cite{54JCP-Green,57JPSJ-Kubo,76-McQuarrie}
where $\kappa_I$ is expressed
by the heat flux auto-correlation function requiring the energy
and the virial tensor of each atom from the molecular dynamics trajectory.
Typically, this formula is used with empirical force fields~\cite{05JCG-Kawamura,12JNR-Taherkhani,17JNM-Kim}
since traditional DFT methods cannot yield explicit energy for each atom;
note that {\it ab initio} GK formulas for evaluation of $\kappa_I$
have been proposed in some recent works.~\cite{17L-Carbogno,16NP-Marcolongo,17B-Kang,19NJP-Martin}
Besides, a long molecular dynamics trajectory may be needed to evaluate $\kappa_I$,
which poses another challenge issue for DFT simulations.

While current experimental techniques only measure the total thermal conductivity,
the first-principles methods have become an ideal tool
to separately yield electronic and ionic contributions;
however, only a few works that adopt first-principles methods
have focused on transport properties by considering both contributions.
~\cite{12AJSS-Martin,16B-Jain,19B-yani}
For some metals, the ionic contribution is not important,~\cite{16B-Jain}
but this observation does not hold for all metals such as tungsten.~\cite{19B-yani}
Besides, the first-principles methods are computationally expensive especially when
a large system or a long trajectory is considered.
In this regard, the community awaits for an efficient and accurate method that can
yield both electronic and ionic thermal conductivities of materials.
Herein, with the above KG and GK formulas,
we propose a combined DFT and deep potential (DP) method for the purpose
and take warm dense Al as an example.
The recently developed DP molecular dynamics (DPMD)
~\cite{18ANIPS-Zhang,18CPC-Wang,18L-Zhang} method
learns first-principles data via deep neural networks
and yields a highly accurate many-body potential to describe the interactions among atoms.
Compared with the traditional DFT, DPMD has a much higher efficiency while
keeps the {\it ab initio} accuracy.
In addition, the linear-scaling DPMD can be parallelized
to simulate hundreds of millions of atoms.~\cite{20ACM-Weile,21CPC-Denghui}
The DP method has been adopted in a variety of applications,
such as crystallization of silicon~\cite{18L-Bonati},
high entropy materials~\cite{20JMST-fuzhi},
isotope effects in liquid water~\cite{19MP-Hsin-Yu,20B-Jianhang},
and warm dense matter~\cite{20JPCM-Qianrui,20PP-Yuzhi}, etc.

In this work, we demonstrate that the DP method in conjunction with DFT can be adopted to obtain
both electronic and ionic thermal conductivities for warm dense Al.
We use KSDFT, OFDFT and DPMD to study the electronic and ionic thermal conductivities
of warm dense Al at temperatures of 0.5, 1.0 and 5.0 eV
with a density of 2.7 ${\rm g/cm^{3}}$.
The electronic thermal conductivity can be accurately computed
via the KG method based on the DPMD trajectories as compared to the FPMD trajectories.
Importantly,  we systematically investigate the convergence issues
such as the number of $k$-points, the number of atoms, the broadening parameter,
the exchange-correlation functionals, and the pseudopotentials in affecting the electronic thermal conductivity with
the aid of DPMD simulations.
%
%We show that the electronic thermal conductivities computed based on pseudopotentials consisting
%of 3 and 11 valence electrons deviate at a temperature of 5.0 eV, which agrees with previous works.
%
Furthermore, the ionic thermal conductivity can be obtained via DPMD and
the convergence is studied with different sizes of systems and different lengths of trajectories.

The manuscript is organized as follows.
In section II, we briefly introduce KSDFT, OFDFT, and DPMD
and the setups of simulations.
We also briefly introduce the Kubo-Greenwood formula and the Green-Kubo formula that used in this work.
In section III, we first show the computed electronic thermal conductivity from the above three methods.
The convergence issues of the electronic thermal conductivity are then thoroughly discussed.
Finally, we show the results of the ionic thermal conductivity of warm dense Al.
We conclude our works in Section IV.

\section{Method}

\subsection{Density Functional Theory}
The ground-state total energy within the formalism of DFT~\cite{64PR-Hohenberg,65PR-Kohn}
can be expressed as a functional dependence of the electron density.
The Hohenberg-Kohn theorems~\cite{64PR-Hohenberg}
point out that the electron density which minimizes the total energy is the ground-state density
while the energy is the ground-state energy.~\cite{64PR-Hohenberg}
According to different treatments with the kinetic energy of electrons,
two methods appear, i.e., KSDFT~\cite{65PR-Kohn} and OFDFT~\cite{02TMCPC-Wang}.
The kinetic energy of electron in KSDFT does not include the electron density
explicitly but is evaluated from the ground-state wave functions of electrons
obtained by the self-consistent iterations.
On the other hand, OFDFT approximately denotes the kinetic energy of electrons
as an explicit density functional~\cite{27MPCPS-Thomas,27RANL-Fermi,28ZP-Fermi,35ZP-vW,92B-Wang},
which enables the direct search for the ground-state density and energy.
As a result, KSDFT has a higher accuracy but OFDFT is several orders more
efficient than KSDFT, especially for large systems.
Besides, Mermin extends DFT to finite temperatures,~\cite{65PR-Mermin}
which has been widely used to describe electrons in WDM.~\cite{02E-Desjarlais,05B-Recoules}

We ran 64-atom Born-Oppenheimer molecular dynamics (BOMD) simulations with KSDFT by using the QUANTUM ESPRESSO 5.4 package.~\cite{17JPCM-Giannozzi}
The Perdew-Burke-Ernzerhof (PBE) exchange-correlation (XC) functional~\cite{96L-PBE}
was used. Projector augmented-wave (PAW) potential~\cite{94B-Blochl,01CPC-Holzwarth} was adopted
with three valence electrons and a cutoff radius of 1.38 \AA.
The plane wave cutoff energy was set to 20 Ry for temperatures of 0.5 and 1.0 eV
and 30 Ry for 5.0 eV. We only used the gamma $k$-point.
Periodic boundary conditions were used.
Besides, the Andersen thermostat~\cite{80JCP-Andersen} was employed with the NVT ensemble
and the trajectory length was 10 ps with a time step of 1.0 fs.

We also performed 108-atom BOMD simulations with OFDFT by utilizing the PROFESS 3.0 package.~\cite{15CPC-Chen}
Both PBE~\cite{96L-PBE} and LDA~\cite{65PR-Kohn}
XC functionals were used, in together with the Wang-Teter (WT)
KEDF~\cite{92B-Wang} and the Nos\'e-Hoover thermostat~\cite{84JCP-Nose,85A-Hoover} in the NVT ensemble.
We ran OFDFT simulations for 10 ps with a time step of 0.25 fs and the energy cutoff was set to 900 eV.
Periodic boundary conditions were utilized.

\subsection{Deep Potential Molecular Dynamics}

The DP method~\cite{18ANIPS-Zhang,18CPC-Wang,18L-Zhang} learns the dependence of the total energy
on the coordinates of atoms in a system and build a deep neural network (DNN) model,
which predicts the potential energy and force of each atom.
In the training process, the total potential energy $E_{tot}$ is decomposed into energies of different atoms:
\begin{equation}
 E_{tot}=\sum_{i} E_i,
\end{equation}
where $E_i$ is the potential energy of atom $i$.
In general, for each atom $i$, the mapping is established through DNN between
$E_i$ and the atomic coordinates of its neighboring atoms within a cutoff radius $r_c$.
Specifically, the DNN model consists of the embedding network and the fitting network.~\cite{18ANIPS-Zhang}
The embedding network imposes constrains to atoms,
enabling atomic coordinates to obey the translation, rotation and permutation symmetries.
On the other hand, the fitting network maps the atomic coordinates from the embedding network to
$E_i$. Next, the training data are prepared as the total energy and the forces acting on atoms
are extracted from the FPMD trajectories.
Finally, the parameters of the DNN model are optimized by
minimizing the loss function. The resulting DNN-based model can
be used to simulate a large system with the efficiency comparable to empirical force fields.

In this work, we adopted the DNN-based models trained from either KSDFT or OFDFT trajectories
with the DeePMD-kit package.~\cite{18CPC-Wang}
With the purpose to study the size effects of electronic thermal conductivity,
a series of cubic cells consisting of 16, 32, 64, 108, 216 and 256 atoms were simulated for 10 ps
with a time step of 0.25 fs.
Besides, we ran systems of 16, 32, 64, 108, 256, 1024, 5488, 8192, 10648, 16384, 32000 and 65536 atoms
in order to investigate the convergence of ionic thermal conductivity.
The length of these trajectories is 500 ps with a time step of 0.25 fs.
We employed the Nos\'e-Hoover thermostat~\cite{84JCP-Nose,85A-Hoover} in the NVT ensemble.
Periodic boundary conditions were adopted.
All of the DPMD simulations were performed by the modified LAMMPS package.~\cite{95JCP-Plimpton}

\subsection{Kubo-Greenwood Formula}

Electronic thermal conductivity $\kappa_e$ is calculated by the Onsager coefficients $L_{mn}$ as
\begin{equation}
\kappa_e=\frac{1}{e^2T}\left(L_{22}-\frac{L_{12}^2}{L_{11}}\right),
\label{kappa_e}
\end{equation}
where $T$ is temperature and $e$ is the charge of electrons,
and $L_{mn}$ is obtained by the frequency-dependent Onsager coefficients $L_{mn}(\omega)$ as
\begin{equation}
	L_{mn}=\lim_{\omega\to0}L_{mn}(\omega).
\end{equation}
We follow the Kubo-Greenwood formula derived by Holst {\it et al.},~\cite{11B-Holst}
and $L_{mn}(\omega)$ takes the form of
\begin{equation}
\begin{aligned}
\label{Ocoe}
&L_{mn}(\omega)=(-1)^{m+n}\frac{2\pi e^2\hbar^2}{3m_e^2\omega\Omega}\\
&\times\sum_{ij\alpha\mathbf{k}}W(\mathbf{k})\left(\frac{\epsilon_{i\mathbf{k}}+\epsilon_{j\mathbf{k}}}{2}-\mu\right)^{m+n-2}|
\langle\Psi_{i\mathbf{k}}|\nabla_\alpha|\Psi_{j\mathbf{k}}\rangle|^2\\
&\times[f(\epsilon_{i\mathbf{k}})-f(\epsilon_{j\mathbf{k}})]\delta(\epsilon_{j\mathbf{k}}-\epsilon_{i\mathbf{k}}-\hbar\omega),
\end{aligned}
\end{equation}	
where $m_e$ is the mass of electrons, $\Omega$ is the volume of cell,
$W(\mathbf{k})$ represents the weight of $\mathbf{k}$ points in the Brillouin zone,
$\mu$ is the chemical potential, $\Psi_{i,\mathbf{k}}$ represents
the wave function of the $i$-th band with the eigenvalue $\epsilon_{i,\mathbf{k}}$
and $f$ being the Fermi-Dirac function.
We used the chemical potential $\mu$ instead of the enthalpy per atom,
which does not affect the results of electronic thermal conductivity in one-component systems.~\cite{11B-Holst}
Note that here we adopted the momentum operator in Eq.~(\ref{Ocoe}) instead of the velocity operator,
which introduces additional approximations due to the use of non-local pseudopotentials.~\cite{91B-Read,07JPCM-Knider,05B-Recoules,12B-Vlcek,17PP-French}
However, the non-local errors were demonstrated to be sufficiently small (about 3.3\%) for liquid Al~\cite{07JPCM-Knider}, which are considered to be reasonably small in this work. 
In future works, we will include the additional term caused by non-local pseudopotentials.
We also define the frequency-dependent electronic thermal conductivity as
\begin{equation}
\kappa_e(\omega)=\frac{1}{e^2T}
\left(L_{22}(\omega)-\frac{L_{12}^2(\omega)}{L_{11}(\omega)}\right).
\label{kappa_e_omega}
\end{equation}
In practical usage of the KG method,
the delta function in Eq.~(\ref{Ocoe}) needs to be broadened.
We adopte a Gaussian function~\cite{02E-Desjarlais}
and the delta function takes the form of
\begin{equation}
\delta(E)=\lim_{\Delta E\to 0}\frac{1}{\sqrt{2\pi}\Delta E}e^{-\frac{E^2}{2{\Delta E}^2}}.
\label{eq:FWHM}
\end{equation}
Here $\Delta E$ controls the full width at half maximum (FWHM, denoted as $\sigma$)
of Gaussian function with the relation of $\sigma\approx2.3548\Delta E$.

\begin{figure}
	\centering
	\includegraphics[width=12cm]{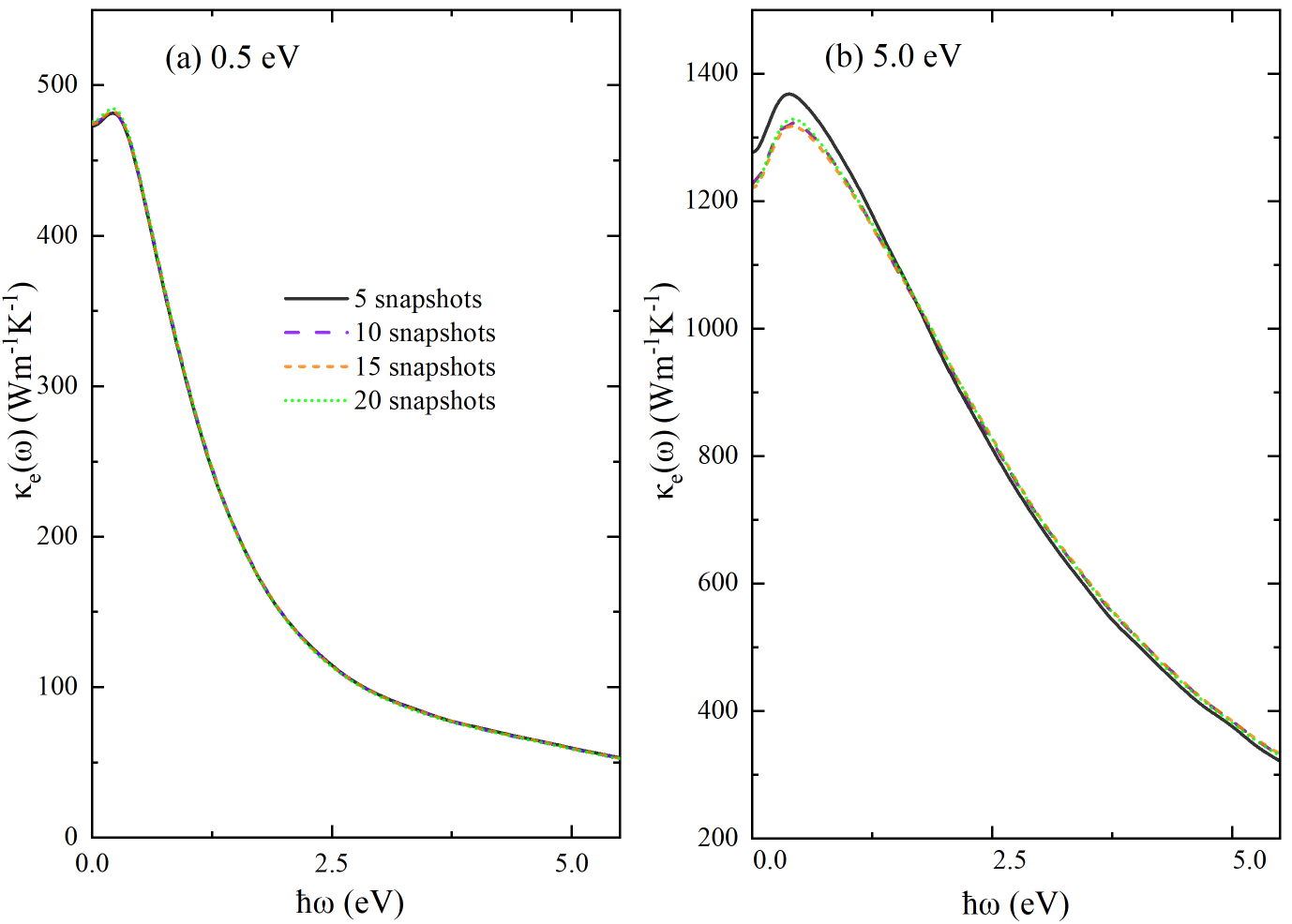}\\
	\caption{
(Color online) Convergence of frequency-dependent electronic thermal conductivity $\kappa_e(\omega)$ of Al with
respect to the number of atomic configurations (snapshots) selected from DPMD trajectories.
The temperatures are set to (a) 0.5 and (b) 5.0 eV.
The number of snapshots used together with the Kubo-Greenwood formula
(the broadening parameter is set to 0.4 eV) is shown with different lines.
The DP model is trained from the OFDFT trajectories using
the PBE exchange-correlation functional. The simulation cell contains 108 atoms.
	}\label{fig:snapshot}
\end{figure}

The KG method needs computed eigenvalues and wave functions from DFT solutions of given atomic configurations.
In practice, we selected 5-20 atomic configurations from the last 2-ps MD trajectories with a time interval of 0.1 ps.
We used both PBE and LDA XC functionals and the associated norm conserving (NC) pseudopotentials in order
to test the influences of XC functionals and pseudopotentials on the resulting electronic thermal conductivity.
We adopted two NC pseudopotentials for Al, which are referred as PP1 and PP2.
The PP1 pseudopotential was generated with the optimized norm-conserving Vanderbilt pseudopotential method
via the ONCVPSP package.~\cite{13B-Hamann,17B-Hamann}
We used 11 valence electrons and a cutoff radius of 0.50 \AA.
Calculations of $\kappa_e$ in a 64-atom cell involved 1770, 2100 and 5625 bands
at temperatures of 0.5, 1.0 and 5.0 eV, respectively.
The PP2 pseudopotential was generated through the PSlibrary package.~\cite{14CMS-Corso}
We used the Troullier-Martins method~\cite{91B-Troullier} and the cutoff radius was set to 1.38~\AA.
We chose 3 valence electrons for each atom. We selected 720, 1100 and 4800 bands
for calculations of $\kappa_e$ at temperatures of 0.5, 1.0 and 5.0 eV, respectively.
The plane wave cutoff energies of both pseudopotentials were set to 50 Ry.
Generally, the PP1 pseudopotential was used in most cases
and the PP2 pseudopotential was only used to compare the effects of different pseudopotentials on $\kappa_e$.
Fig.~\ref{fig:snapshot} shows the computed averaged $\kappa_e(\omega)$
with respect to different numbers of atomic configurations from 108-atom DPMD trajectories at 0.5 and 5.0 eV.
The DP model was trained based on the 108-atom OFDFT trajectories using
the PBE XC functional.
We can see that 20 atomic configurations are enough to converge $\kappa_e(\omega)$.
Additionally, $\kappa_e(\omega)$ is easier to converge at the relatively lower temperature of 0.5 eV.
Therefore, we chose 20 atomic configurations for cells with 108 atoms or less at all of the three temperatures.
For systems with the number of atoms larger than 108, we respectively selected 5 and 10 atomic configurations
for temperatures smaller than 5.0 eV and equal to 5.0 eV unless otherwise specified.

\subsection{Green-Kubo Formula}
In the DPMD method, the total potential energy of the system
is decomposed onto each atom.
In this regard, the ionic thermal conductivity
can be calculated through the GK method~\cite{76-McQuarrie} with the formula of
\begin{equation}
    \label{ki}
	\kappa_I=\frac{1}{3\Omega k_BT^2}\int_0^{+\infty}\langle \mathbf{J_q}(t)\cdot \mathbf{J_q}(0)\rangle dt,
\end{equation}
where $\Omega$ is the volume of cell, $T$ is temperature,
$k_B$ is the Boltzmann constant,
$\langle\cdots\rangle$ is ensemble average and $t$ is time.
$\mathbf{J_q}$ in Eq.~(\ref{ki}) is the heat current of a one-component system in the center-of-mass frame which takes the form of
\begin{equation}
 \label{heat_flux}
 \mathbf{J_q}=\sum_{i=1}^{N}\varepsilon_i
 \mathbf{v_i}-\frac{1}{2}\sum_{i=1}^N\sum_{j\ne i}^N(\mathbf{v_i}\cdot\mathbf{F}_{ij})\mathbf{r}_{ij}.
\end{equation}
Here $\mathbf{v_i}$ is the velocity of the {\it i}-th atom,
while $\varepsilon_i$ is the energy of the {\it i}-th atom including ionic kinetic energy and potential energy, where the potential energy is directly obtained by DNN.
$\mathbf{F}_{ij}$ is the force acting on the {\it i}-th atom due to the presence of the {\it j}-th atom with
$\mathbf{r}_{ij}$ defined as $\mathbf{r}_i-\mathbf{r}_j$.
It is worth mentioning that Eq.~(\ref{heat_flux}) is valid for 2-body potentials but yields
about 20\% error for $\mathbf{J_q}$ when a many-body potential (e.g. the deep potential) is adopted~\cite{19JCTC-Boone}.
Importantly, as will be shown below, the ionic thermal conductivity of warm dense Al is at least
two orders of magnitudes smaller than the electronic thermal conductivity counterpart.
Therefore, we still consider Eq.~(\ref{heat_flux}) as a valid but approximated formula
to yield the ionic thermal conductivity of warm dense Al with DPMD. Additionally, we recommend the usage of a
more complete formula for those systems whose ionic thermal conductivity is an important part of the
total thermal conductivity.

Eq.~(\ref{ki}) can also be written as
\begin{equation}
\label{kappa_i}
\kappa_I=\int_0^{+\infty}C_J(t) dt,
\end{equation}
from which the auto-correlation function of heat current is defined as
\begin{equation}
C_J(t)=\frac{1}{3\Omega k_BT^2}\langle \mathbf{J_q}(t)\cdot \mathbf{J_q}(0)\rangle.
\label{eq:Ct}
\end{equation}

%Note that the ionic thermal conductivity of Al is only computed via
%DPMD simulations since no single atom energy is available from traditional DFT calculations.

\section{Results}
\subsection{Accuracy of DP models}

We first performed FPMD simulations of Al based on KSDFT and OFDFT at temperatures of 0.5, 1.0 and 5.0 eV.
The PBE XC functional was used and the FPMD trajectory length was 10 ps. The cell contained 64 Al atoms.
Two DP models named DP-KS and DP-OF were trained based on the KSDFT and OFDFT trajectories, respectively.
Note that the accuracy of the DP models in describing warm dense Al at the three temperatures have
been demonstrated in our previous work,~\cite{20JPCM-Qianrui}
where we show that DPMD has an excellent accuracy in reproducing structural and dynamical properties
including radial distribution functions, static structure factors, and dynamic structure factors.

\begin{figure}
	\centering
	\includegraphics[width=12cm]{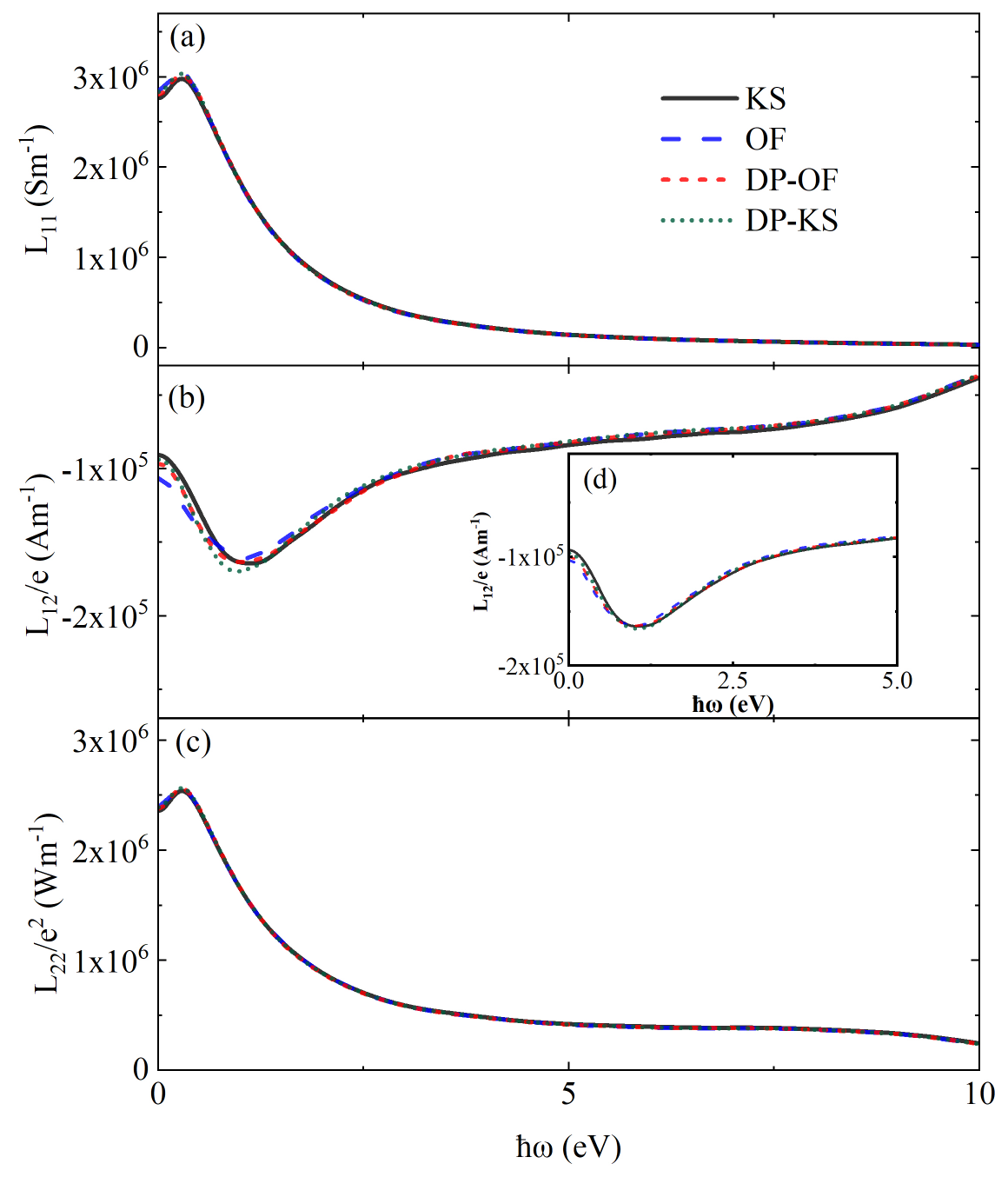}\\
	\caption{(Color online) Frequency-dependent Onsager kinetic coefficients (a) $L_{11}$, (b) $L_{12}$, and (c) $L_{22}$
of Al at a temperature of 0.5 eV as computed from KS, OF, DP-KS and DP-OF methods.
DP-KS and DP-OF refer to the DP models trained from OFDFT and KSDFT molecular dynamics trajectories, respectively.
The broadening parameter used in the Kubo-Greenwood method is set to 0.4 eV.
The simulation cell contains 64 Al atoms.
	}\label{fig:Onsager}
\end{figure}

Here, we first focus on the frequency-dependent Onsager coefficients $L_{mn}(\omega)$.
Fig.~\ref{fig:Onsager} shows the computed $L_{11}(\omega)$,
$L_{12}(\omega)$, and $L_{22}(\omega)$ via different methods
at 0.5 eV;
$L_{21}(\omega)$ is not illustrated since $L_{21}(\omega)=L_{12}(\omega)$ as derived from Eq.~(\ref{Ocoe}).
We can see that all of the four methods yield very close $L_{11}(\omega)$ and $L_{22}(\omega)$
except for $L_{12}(\omega)$, where slightly larger differences are found at low frequencies.
The main reason is that $L_{12}(\omega)$ is more sensitive to the number of
ionic configurations and 20 snapshots are not enough to converge it well.
We also test 40 snapshots and the results are improved, as illustrated in Fig.~\ref{fig:Onsager}(d).
However, the resulting electronic thermal conductivity mainly depends on $L_{22}(\omega)$
at temperatures considered in this work and
the small differences of $L_{12}(\omega)$ do not affect the computed $\kappa_{e}(\omega)$,
as will be shown next.

\begin{figure}
	\centering
	\includegraphics[width=12cm]{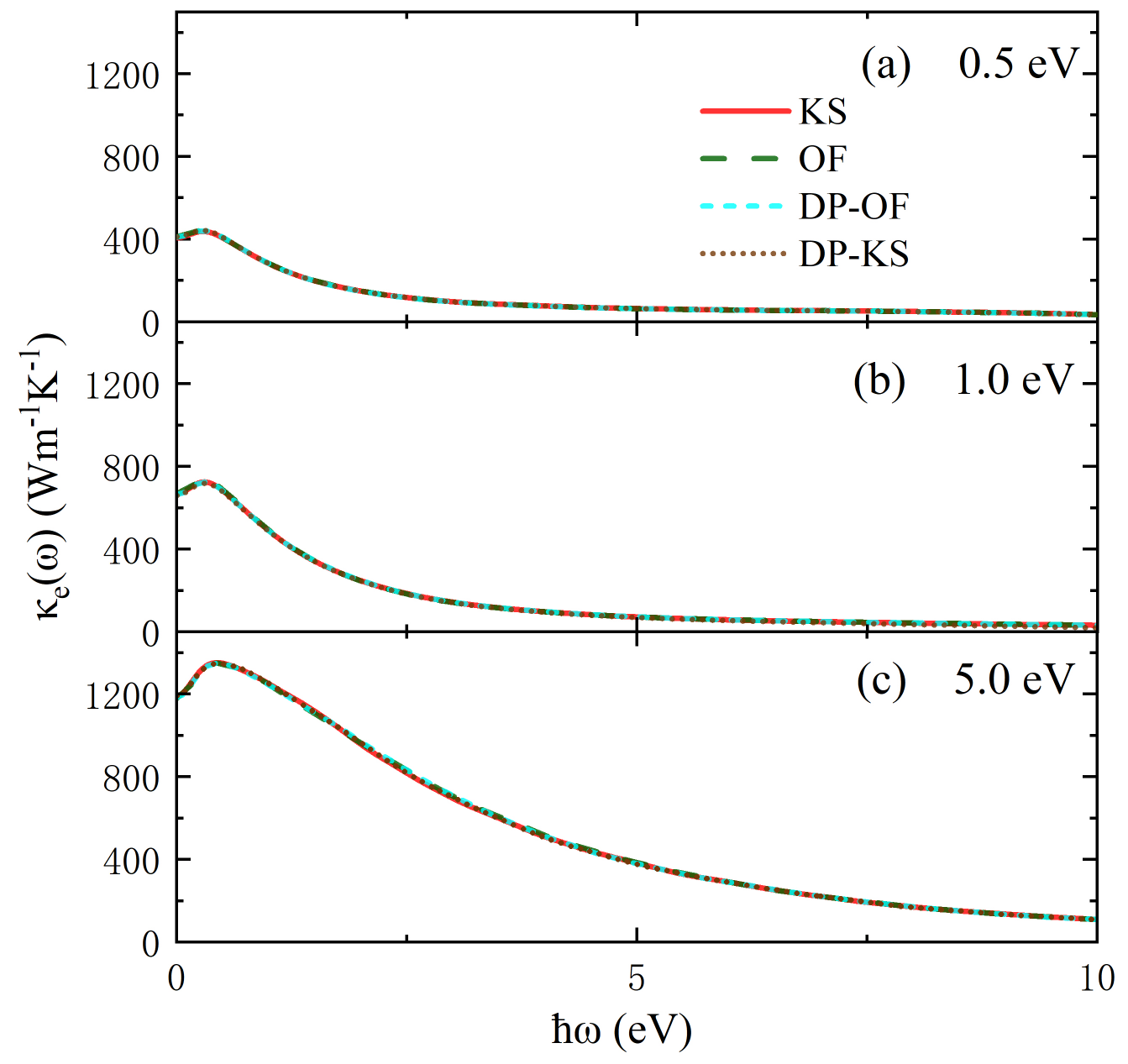}\\
	\caption{(Color online) Frequency-dependent electronic thermal conductivity $\kappa_{e}(\omega)$
as computed from the Kubo-Greenwood method (the broadening parameter is set to 0.4 eV)
with snapshots from KS, OF, DP-KS and DP-OF molecular dynamics trajectories.
The temperatures are (a) 0.5, (b) 1.0 and (c) 5.0 eV.
DP-KS and DP-OF refer to the DP models trained from OFDFT and KSDFT molecular dynamics trajectories, respectively.
The cell contains 64 Al atoms.
	}\label{fig:compare}
\end{figure}

The computed frequency-dependent electronic thermal conductivity
$\kappa_e(\omega)$ utilizing the abovementioned four different methods
are illustrated in Fig.~\ref{fig:compare}.
We can see that the OFDFT results agree well with the KSDFT ones,
suggesting OFDFT has the same accuracy as KSDFT to yield
atomic configurations for subsequent computations of $\kappa_e(\omega)$ using the KG method,
even though there are some approximations on the kinetic energy of electrons~\cite{92B-Wang}
and the local pseudopotential~\cite{08PCCP-Huang} within the framework of OFDFT.
Impressively, the DP models trained from FPMD trajectories yield
almost identical $\kappa_e(\omega)$ when compared to the DFT results,
which proves that the DP models can yield highly accurate atomic configurations
for subsequent calculations of $\kappa_e(\omega)$.
In conclusion, the input atomic configurations for the calculation of $\kappa_e(\omega)$
can be generated by efficient DPMD models without losing accuracy,
which is beneficial for simulating a large number of atoms to mitigate size effects.
Although the preparation of the training data in the DPMD model requires
additional computational resources,
we find running FPMD with a cell consisting of 64 atoms is sufficient to generate reliable DPMD models.
Therefore, additional computational costs are saved
once larger numbers of atoms are adopted in the linear-scaling DPMD method.~\cite{20JPCM-Qianrui}

\subsection{Convergence of Electronic Thermal Conductivity}

Previous works~\cite{11B-Pozzo,11PP-Flavien,13CMS-Knyazev} have shown that the electronic thermal conductivity $\kappa_{e}$
depends strongly on both the number of $k$-points and the number of atoms.
Additionally, the broadening parameter should be properly chosen in order to yield a meaningful $\kappa_{e}$.
Herein, we systematically investigate the above issues by
adopting the DPMD model to generate atomic configurations for
subsequent calculations of $\kappa_{e}$.
The DPMD model was trained from OFDFT-based MD trajectory with the PBE XC functional.
By utilizing the snapshots from the DPMD trajectory,
we obtain $\kappa_{e}$ by using the KG formula.
Furthermore, we study the effects of different
XC functionals and pseudopotentials in affecting $\kappa_{e}$.

\subsubsection{Number of $k$-points}

\begin{figure}
	\centering
	\includegraphics[width=12cm]{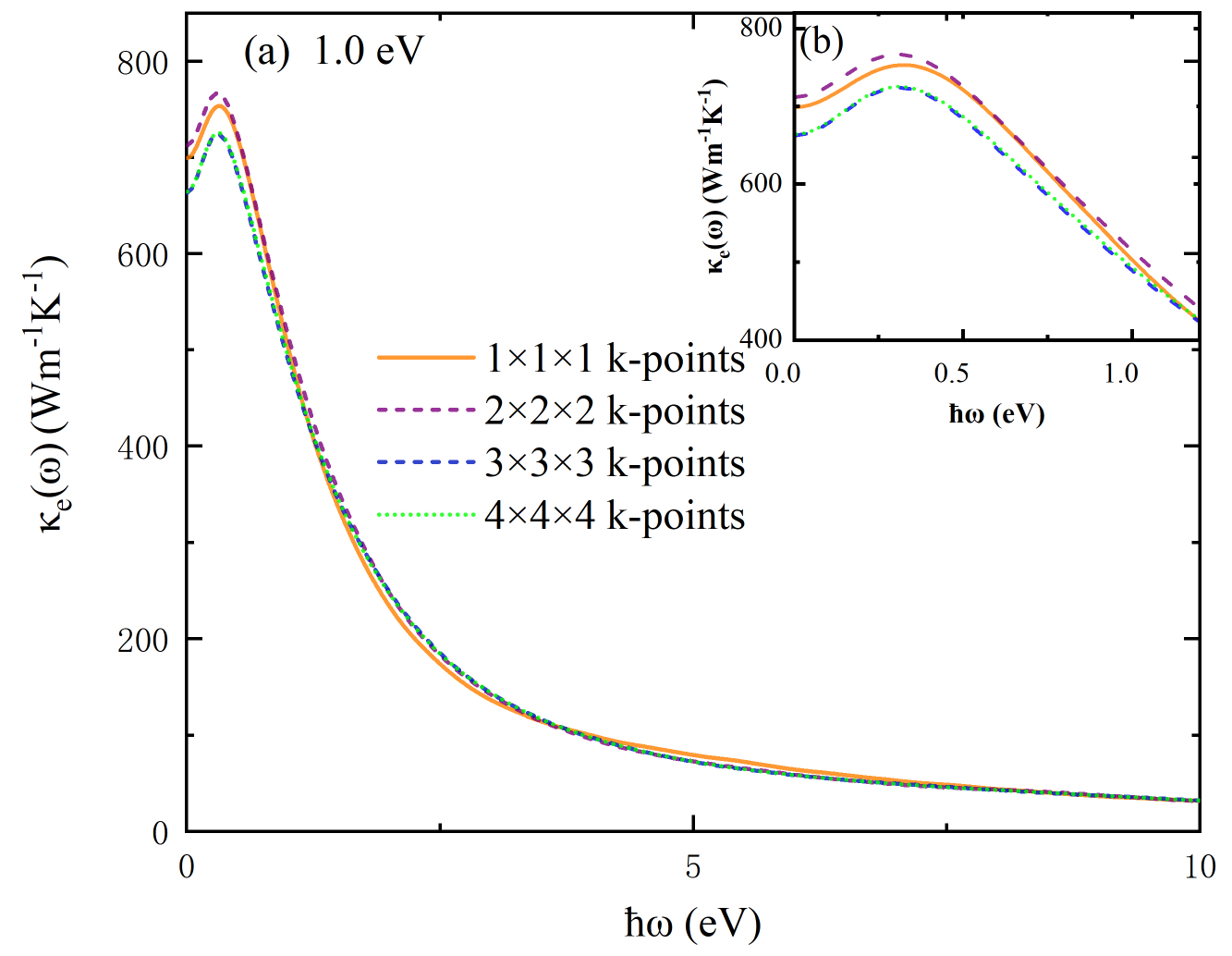}
	\caption{(Color online) Convergence of frequency-dependent electronic thermal conductivity $\kappa_{e}(\omega)$ of Al
with respect to the number of $k$-points. The temperature is set to 1.0 eV.
The $k$-point samplings utilized with the Kubo-Greenwood method (the broadening parameter is set to 0.4 eV)
are chosen from 1$\times$1$\times$1 to 4$\times$4$\times$4. The DP model is trained from the OFDFT trajectories using the PBE exchange-correlation functional.
The cell consists of 64 atoms.
	}\label{fig:k-conv}
\end{figure}

We first investigate the convergence of $\kappa_e(\omega)$
with respect to different $k$-points.
Fig.~\ref{fig:k-conv} illustrates an example of warm dense Al in a 64-atom cell at a temperature of 1.0 eV.
The $k$-point meshes were chosen to be $1\times1\times1$, $2\times2\times2$,
$3\times3\times3$, and $4\times4\times4$ in calculations of $\kappa_e(\omega)$.
The results show that $\kappa_e(\omega)$ converges when $3\times3\times3$ $k$-points are used.
In fact, the required number of $k$-points for convergence of $\kappa_{e}(\omega)$
varies with different number of atoms and different temperatures.
Table~\ref{tab:K-points} lists the sizes of $k$-points that are needed to converge
$\kappa_e(\omega)$ with numbers of atoms ranging from 16 to 256 at temperatures of 0.5, 1.0 and 5.0 eV.
We find that the needed number of $k$-points exhibits a trend to
decrease at higher temperatures.
For instance, the needed $k$-point samplings of a 32-atom cell at temperatures of 0.5, 1.0 and 5.0 eV
are $6\times6\times6$, $5\times5\times5$ and $3\times3\times3$, respectively.
This can be understood by the fact that the Fermi-Dirac distribution of electrons
around the Fermi energy becomes more sharp at low temperatures.
Therefore, a dense mesh of $k$-points is needed to represent the detailed
structures around the Fermi surface at low temperatures.

\subsubsection{Number of atoms}

\begin{figure}
	\centering
	\includegraphics[width=12cm]{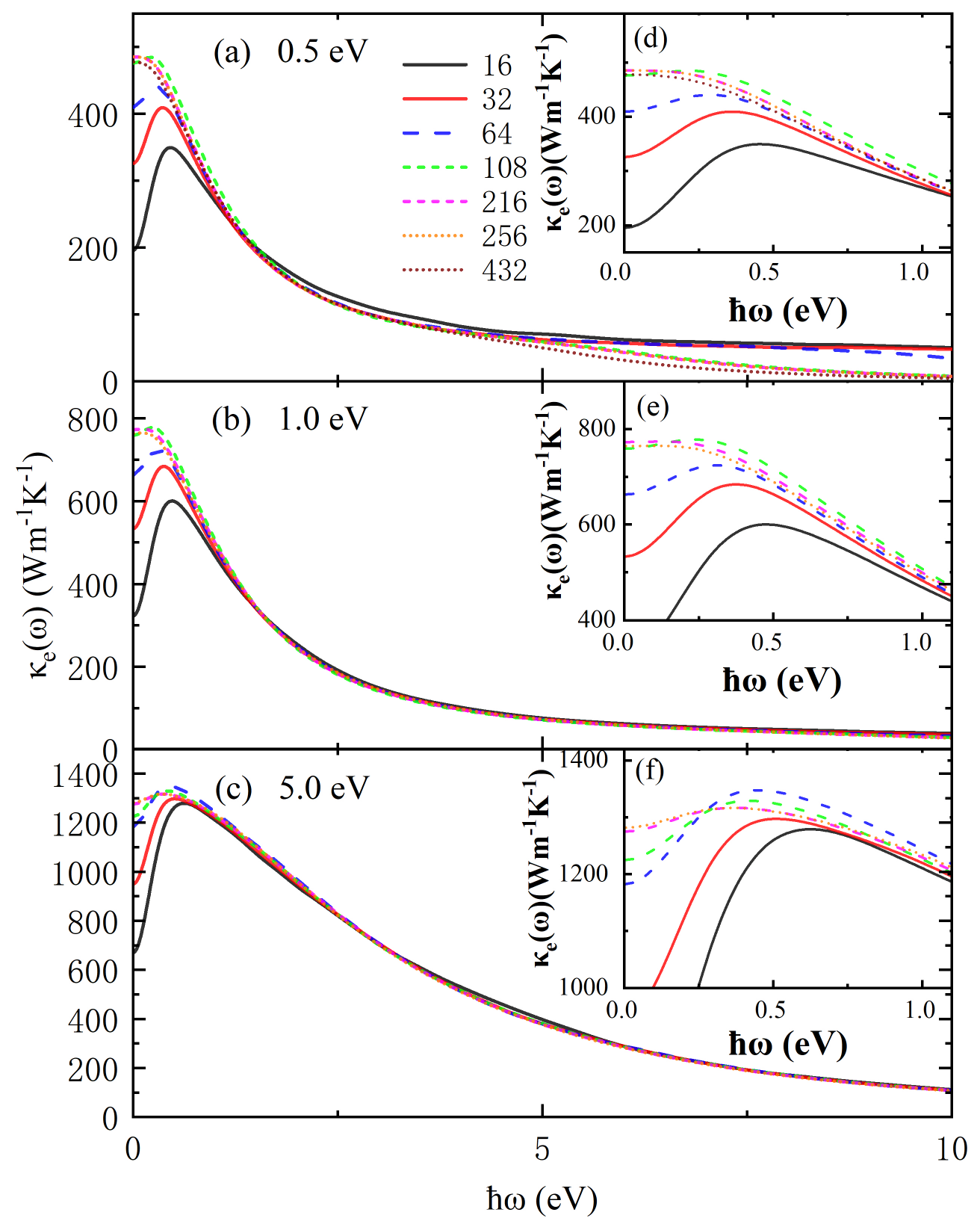}
	\caption{(Color online) Convergence of the electronic thermal conductivity $\kappa_e(\omega)$
 with respect to the number of atoms (from 16 to 432 atoms) in the simulation cell.
 The temperatures are set to (a) 0.5, (b) 1.0 and (c) 5.0 eV. (d-f) illustrate the peaks in (a-c), respectively.
 For the temperature of 0.5 eV, two 432-atom snapshots with $2\times2\times2$ $k$-points are chosen to
 test the convergence of $\kappa_e(\omega)$. The DP model is trained from the OFDFT trajectories using the PBE exchange-correlation functional. The broadening parameter is set to 0.4 eV.
	}\label{fig:conv}
\end{figure}

The value of $\kappa_e(\omega)$ converges not only with enough number of $k$-points but also
with sufficient number of atoms in the simulation cell.
To demonstrate this point, we plot $\kappa_e(\omega)$ with respect to different numbers of atoms in Fig.~\ref{fig:conv}.
For each size of cell, the number of $k$-points is chosen large enough to converge $\kappa_e(\omega)$,
as listed in Table~\ref{tab:K-points}.
We have the following findings.
First, for the temperatures of 0.5, 1.0 and 5.0 eV, we find that a 256-atom system
is large enough to converge $\kappa_e(\omega)$.
Second, most of $\kappa_e(\omega)$ do not monotonically decrease but
have peaks at low frequencies ranging from 0.3 to 0.6 eV.
These peaks move towards $\omega$=0 when a larger number of atoms are utilized,
and almost disappear in large cells such as the 432-atom cell at 0.5 eV.
Similar to previous works~\cite{11B-Pozzo}, this phenomenon is caused by the size effects.
Third, $\kappa_e(\omega)$ converges faster with respect to the number of atoms at high temperatures,
suggesting that the size effects are less significant at higher temperatures.
For example, $\kappa_e$ of the 16- and 64-atom systems
at 0.5 eV are 59.7\% (195.5 ${\rm Wm^{-1}K^{-1}}$) and 15.6\% (409.3 ${\rm Wm^{-1}K^{-1}}$)
lower than the value from a 256-atom system (485.1 ${\rm Wm^{-1}K^{-1}}$).
Meanwhile, $\kappa_e$ of the 16- and 64-atom systems at 5.0 eV are only 47.7\% (669.7 ${\rm Wm^{-1}K^{-1}}$) and 7.5\% (1182.2 ${\rm Wm^{-1}K^{-1}}$) lower than the value from a 256-atom system (1281.7 ${\rm Wm^{-1}K^{-1}}$).

\begin{figure}
	\centering
	\includegraphics[width=12cm]{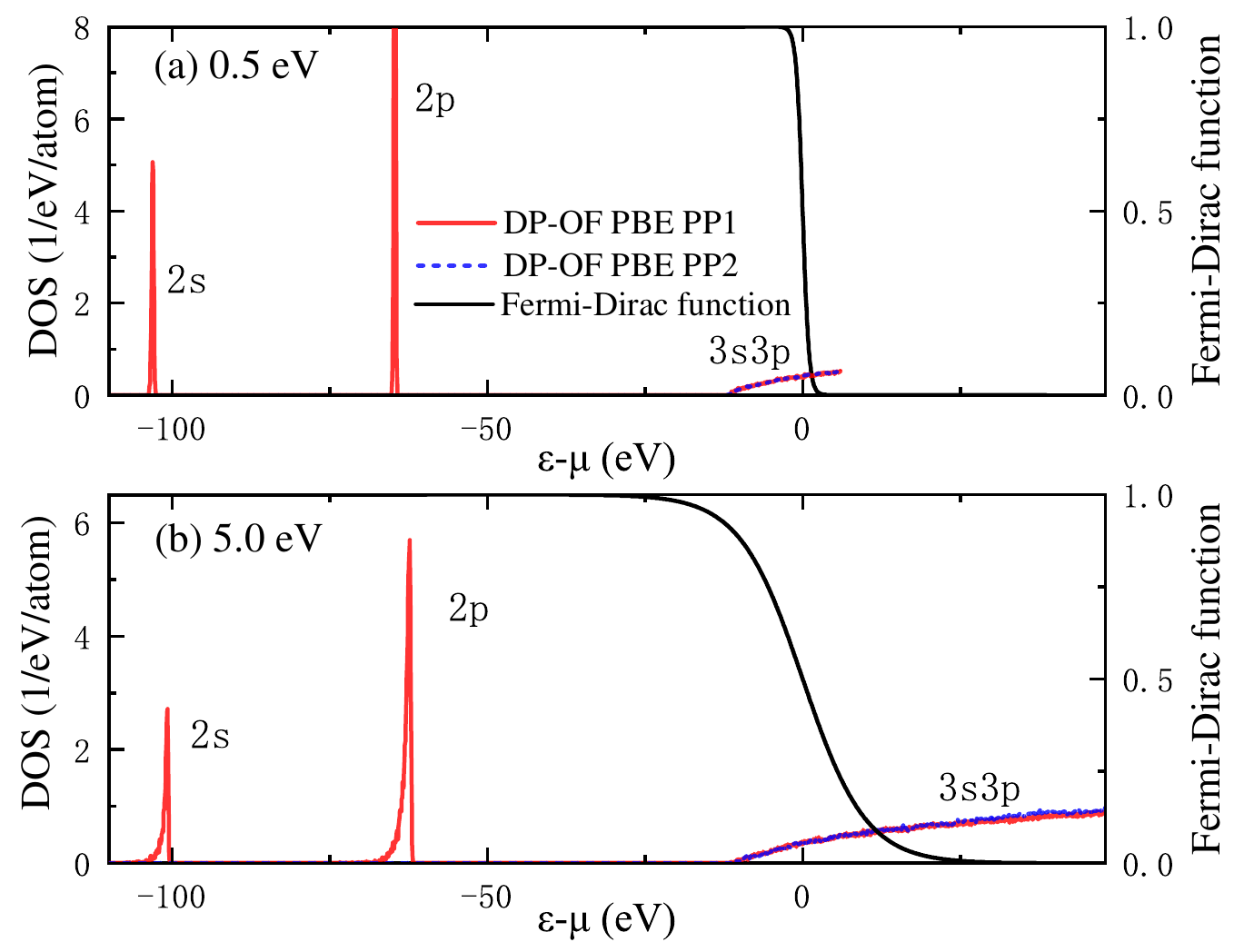}
	\caption{(Color online) Density of states of a 256-atom cell at temperatures of (a) 0.5 and (b) 5.0 eV.
The Fermi-Dirac function at the same temperature is plotted with a black solid line. The DP-OF model
refers to the DP model trained from OFDFT molecular dynamics trajectory.
	}\label{fig:dos}
\end{figure}

\begin{figure}
	\centering
	\includegraphics[width=12cm]{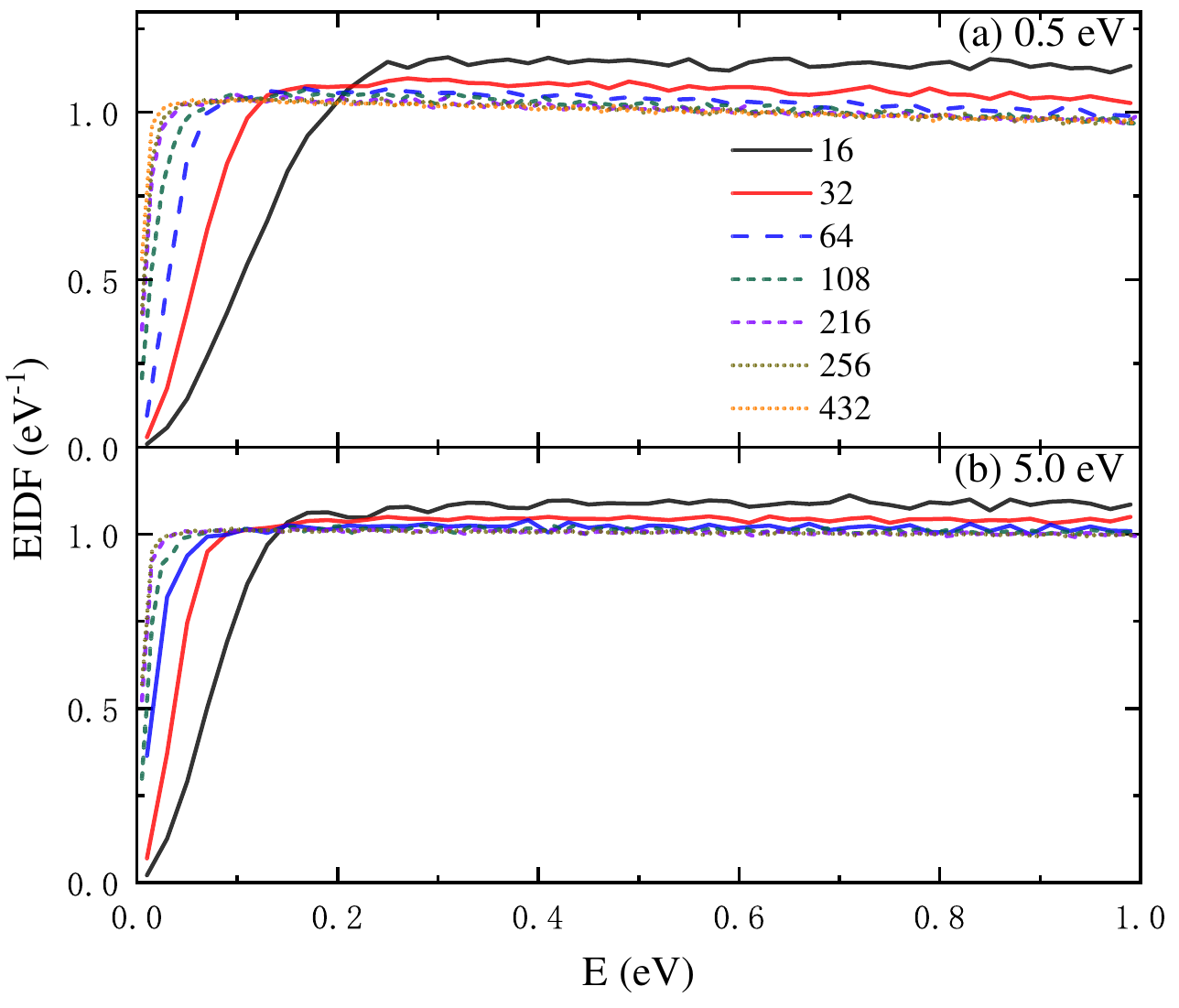}
	\caption{(Color online) Energy interval distribution function of different cells at (a) 0.5 eV and (b) 5.0 eV.
The bands within 6.0 and 50.5 eV above the chemical potential $\mu$ are considered
for temperatures of 0.5 and 5 eV, respectively.
Different lines refer to different numbers of atoms (from 16 to 432 atoms) in the simulation cell.
The DP model is trained from the OFDFT trajectories using the PBE exchange-correlation functional.
	}\label{fig:a-eidf}
\end{figure}

We perform further analysis to elucidate the origin of size effects in computations of $\kappa_e(\omega)$,
which is due to insufficient small energy intervals caused by limited sizes of systems.
As Eq.~(\ref{Ocoe}) shows, for a given energy interval $\epsilon_{i\mathbf{k}}-\epsilon_{j\mathbf{k}}$ with electronic
states of $i$ and $j$ at a specific $k$ point,
$\kappa_e(\omega)$ should converge with enough electronic eigenstates;
however, Fig.~\ref{fig:conv} illustrates that the computed $\kappa_e(\omega)$ substantially becomes larger
at low frequencies with increased number of atoms in the simulation cell.
The results imply that the information of small energy intervals gained from finite-size DFT calculations
are insufficient to evaluate $\kappa_e(\omega)$
even when the number of $k$-points reaches convergence but a small number of atoms is adopted.
To clarify this issue, we define an energy interval distribution function (EIDF) as
\begin{equation}
   g(E)=\frac{1}{N_p}\sum_{i>j,\mathbf{k}}W(\mathbf{k})\delta(\epsilon_{i\mathbf{k}}-\epsilon_{j\mathbf{k}}-E),
\end{equation}
where $W(\mathbf{k})$ represents the weights of $k$-points used in DFT calculations,
$\epsilon_{i\mathbf{k}}$ and $\epsilon_{j\mathbf{k}}$ are the eigenvalues
and $N_p$ is a normalization factor.
We chose warm dense Al systems at temperatures of 0.5 and 5.0 eV
with the selected energy intervals computed from the bands occupied by 3$s$ and 3$p$ electrons,
which can be identified from density of states (DOS), as illustrated in Fig.~\ref{fig:dos}.
The energy intervals satisfy the condition that $E<$1.0 eV.
In addition, we consider the bands within 6.0 and 50.5 eV above the chemical potential $\mu$
for temperatures of 0.5 and 5 eV, respectively.
The results of $g(E)$ with respect to different numbers of atoms ranging from 16 to 432 atoms
are illustrated in Fig.~\ref{fig:a-eidf}.
The EIDF $g(E)$ becomes larger for small $E$ as the number of atoms increases and
converges when the number of atoms reaches 256 for both cases at 0.5 and 5.0 eV.
The increase of $g(E)$ at small $E$ implies that more energy eigenvalues that have close values
appear in a larger cell with more atoms. In other words,
because of the finite number of atoms used in the simulation cell,
the energy levels are discretized to some extent, resulting in
the lack of energy intervals especially for those small values.
Particularly, these small energy intervals are of significant importance
in evaluating $\kappa_{e}(\omega)$ when $\omega\rightarrow0$ as shown in Eq.~(\ref{Ocoe}).

\subsubsection{Broadening Parameter}

\begin{figure}
	\centering
	\includegraphics[width=12cm]{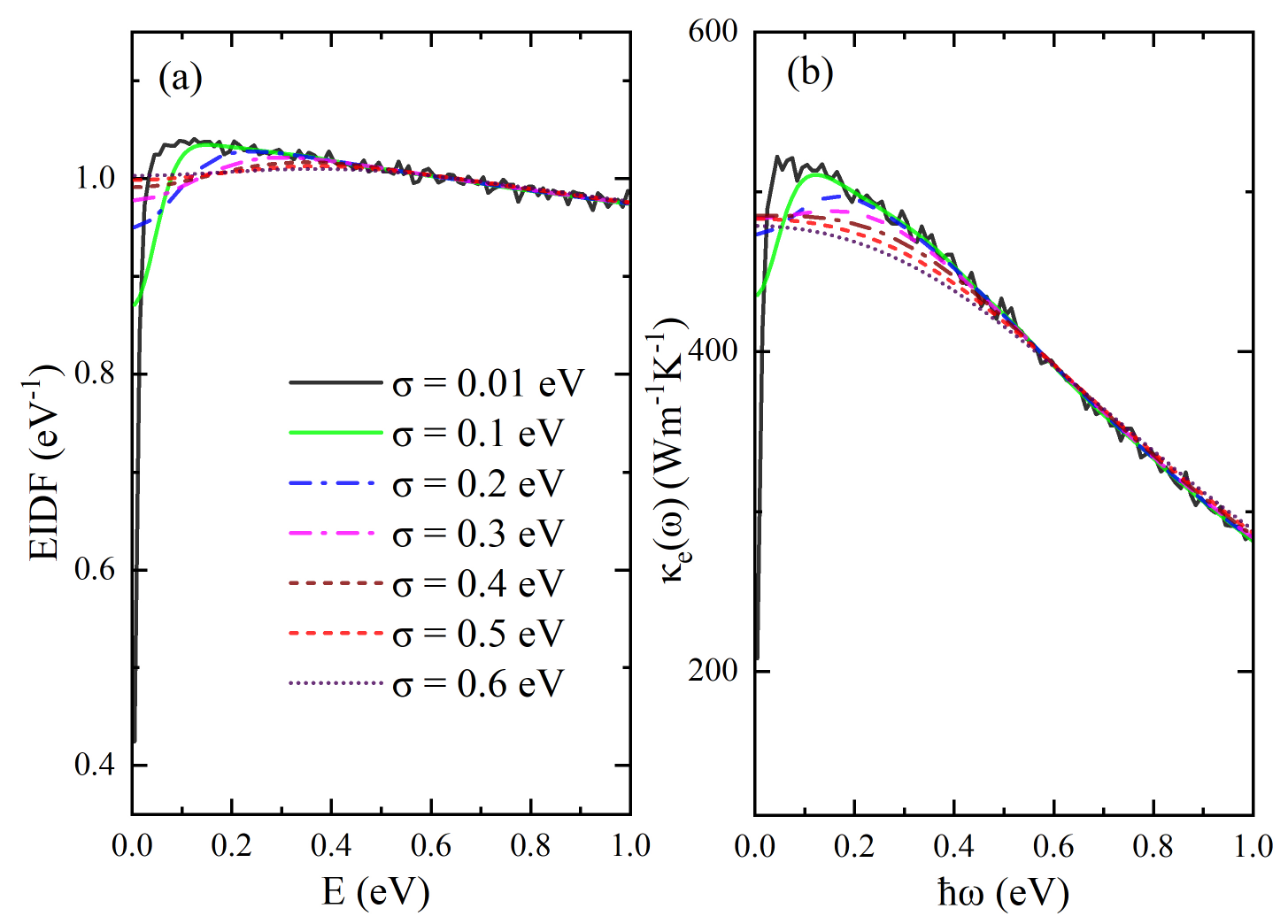}\\
	\caption{(Color online)  (a) Energy interval distribution function and (b) electronic thermal conductivities of a 256-atom cell at 0.5 eV. The snapshots are from DPMD simulations. The DPMD model is trained from OFDFT trajectories with the PBE exchange-correlation functional.
Different lines indicate different broadening parameter $\sigma$.
	}\label{fig:fwhm}
\end{figure}

\begin{figure}
	\centering
	\includegraphics[width=12cm]{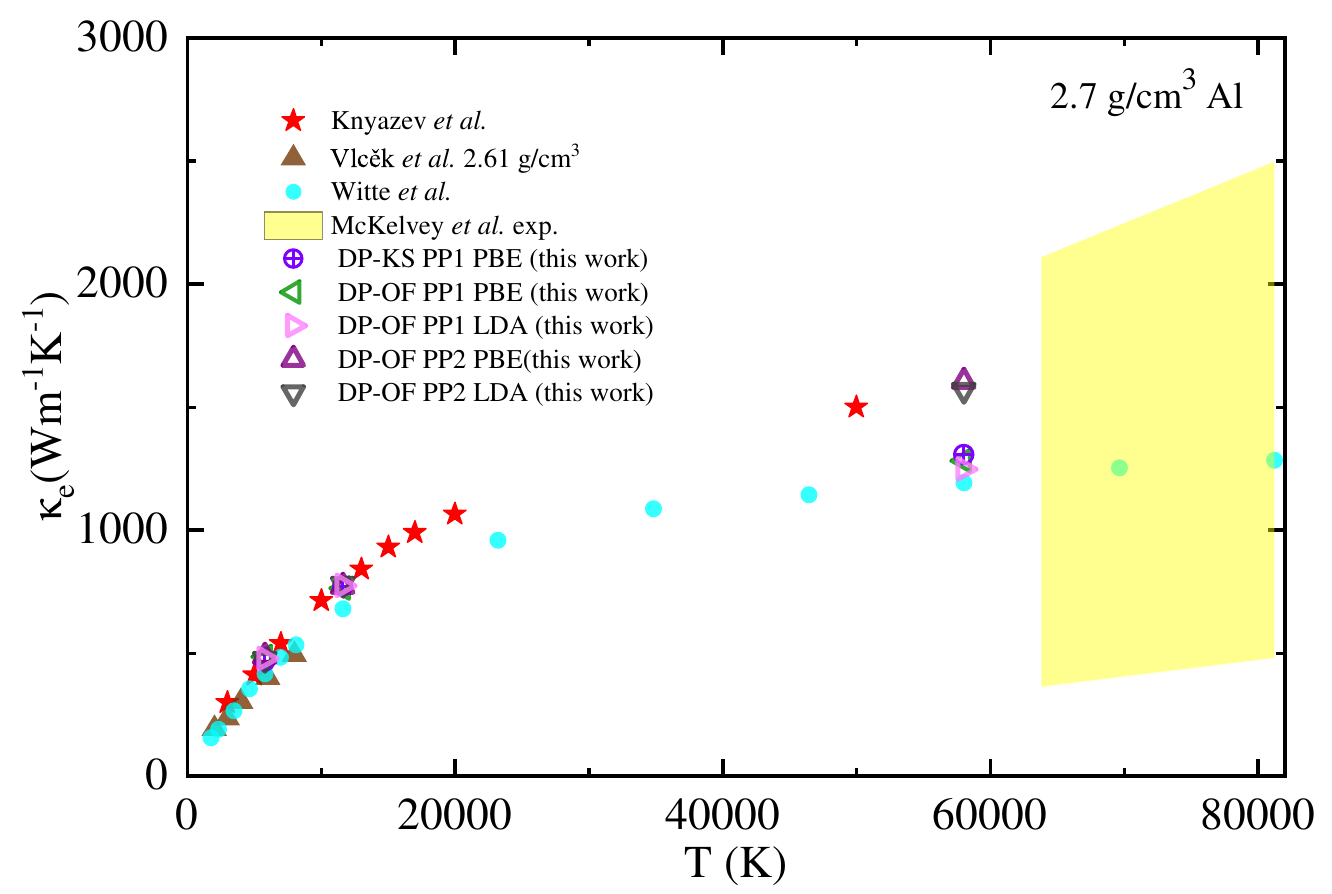}\\
	\caption{
(Color online) Electronic thermal conductivities $\kappa_{e}$ of warm dense Al.
DP-KS and DP-OF refer to the DPMD models that trained from KSDFT and OFDFT, respectively.
Atomic configurations are generated from DP-KS and DP-OF models.
The broadening parameter is set to 0.4 eV.
Note that non-local corrections have been neglected in this study.
Results from Knyazev {\it et al.}~\cite{14PP-Knyazev}, Vlc\v{e}k {\it et al.}~\cite{12B-Vlcek}
and Witte {\it et al.}~\cite{18PP-Witte}, as well as the experimental results from McKelvey {\it et al.}~\cite{17SR-McKelvey}
are shown for comparison.
}\label{fig:kappa}
\end{figure}

The FWHM broadening parameter $\sigma$ that appears in the $\delta(E)$ function in Eq.~(\ref{eq:FWHM})
substantially affects the resulting electronic thermal conductivity $\kappa_e(\omega)$ when $\omega\rightarrow0$.
We therefore investigate the choices of $\sigma$ in influencing the computed $\kappa_e(\omega)$ by analyzing
the EIDF $g(E)$.
Taking Al at 0.5 eV as an example, we plot in Fig.~\ref{fig:fwhm} both $g(E)$ and $\kappa_e(\omega)$ of a 256-atom cell,
which is large enough to converge $g(E)$ or $\kappa_e(\omega)$ as demonstrated above.
When the broadening effect is small ($\sigma$=0.01 eV),
$g(E)$ and $\kappa_e(\omega)$ decrease dramatically below 0.2 eV,
which is caused by the discrete band energies.
Thus, a suitable $\sigma$ needs to be chosen to compensate the discrete band energies.
For instance, the value of $g(E)$ at $E$=0 increases from 0.22 $\rm eV^{-1}$ ($\sigma$=0.01 eV) to 0.99 $\rm eV^{-1}$ ($\sigma$=0.4 eV).
However, $g(E)$ becomes saturated if a too large $\sigma$ is applied, resulting
in overcorrected $\kappa_e(\omega)$ that the curve at frequencies lower than 0.6 eV decreases.
Therefore, in order to compensate the discrete band energies and avoid overcorrection at the same time,
we choose $\sigma$ to be 0.4 eV for warm dense Al at all of the temperatures considered in this work.
It is worth mentioning that a majority of works 
including this work treat the broadening parameter as an adjustable variable 
to yield a better characterized line of frequency-dependent thermal conductivity.~\cite{02E-Desjarlais,13CMS-Knyazev,14PP-Knyazev,05B-Recoules,11PP-Flavien,12B-Vlcek,11B-Pozzo} 
Even though some physical criteria have been proposed~\cite{19PP-Kang}, 
the choice of the adjustable variable is still inconclusive.
In this work, we choose the adjustable parameter in terms of sufficient 
small energy intervals, which are possible when a sufficiently large supercell is used in DFT calculations.
The $\kappa_e$ at zero frequency is obtained by the linear extrapolation
and illustrated in Fig.~\ref{fig:kappa}.

\subsubsection{Exchange-correlation functionals}

We study the influences of LDA and PBE XC functionals on the computed $\kappa_e(\omega)$
by first validating the atomic configurations generated by FPMD simulations.
Specifically, atomic configurations were chosen from two 256-atom DPMD trajectories,
which were generated by two DP models trained from OFDFT with the LDA and PBE XC functionals.
We then adopted the KG method by using the PP1 pseudopotentials
generated with the same XC functional, and yielded $\kappa_e$ at temperatures of 0.5, 1.0, and 5.0 eV.
As shown in Fig.~\ref{fig:kappa} and listed in Table.~\ref{tab:kappa},
the $\kappa_e$ values obtained from the PBE XC functional
are 485.1, 764.1 and 1281.7 ${\rm Wm^{-1}K^{-1}}$
at temperatures of 0.5, 1.0 and 5.0 eV, respectively, while
the $\kappa_e$ values from the LDA XC functional are 1.6\% lower (477.3 ${\rm Wm^{-1}K^{-1}}$),
1.3\% higher (773.8 ${\rm Wm^{-1}K^{-1}}$) and 2.8\% lower (1246.4 ${\rm Wm^{-1}K^{-1}}$)
than those of PBE at 0.5, 1.0 and 5.0 eV respectively.
Therefore, we conclude that the LDA XC functional yields almost the same $\kappa_e$ values as PBE.
Note that a recent work~\cite{18PP-Witte} adopted the HSE XC functional
and showed some differences in electronic thermal and electrical conductivities of warm dense Al
as compared with those from the PBE XC functional.
%
%Note that a previous work suggests
%that the hybrid functional largely affects $\kappa_e$ as compared to PBE.}~\cite{17L-Witte}

\subsubsection{Pseduopotentials}

We investigate how norm-conserving pseudopotentials affect the computed electronic thermal conductivity $\kappa_e$.
First of all, Fig.~\ref{fig:dos} shows DOS of two types of pseudopotentials (PP1 and PP2)
at temperatures of 0.5 and 5.0 eV, where
we see that the two pseudopotentials yield similar DOS of 3$s$3$p$ electrons.
Next, Fig.~\ref{fig:kappa} illustrates the computed $\kappa_e$ from two types of pseudopotentials and
those computed data from Knyazev {\it et al.}~\cite{14PP-Knyazev},
Vlc\v{e}k {\it et al.}~\cite{12B-Vlcek} and Witte {\it et al.}~\cite{18PP-Witte},
as well as the experimental data from McKelvey {\it et al.}~\cite{17SR-McKelvey}.
Besides, $\kappa_e$ from two types of pseudopotentials are
also listed in Table.\ref{tab:kappa}.

We have the following findings.
First, the DP-OF results agree reasonably well with the DP-KS ones, as have been previously shown in Fig.~\ref{fig:compare}.
For example, DP-KS with the PP1 pseudopotential and the PBE XC functional yields $\kappa_e$=466.5 ${\rm Wm^{-1}K^{-1}}$ at 0.5 eV,
which is 3.8\% lower than that of DP-OF (485.1 ${\rm Wm^{-1}K^{-1}}$) at the same temperature and the relative difference decreases to 1.9\% at 5.0 eV.
The above results imply that the OFDFT is suitable to study
the electronic thermal conductivity of warm dense Al ranging from 0.5 to 5.0 eV.
Second, our calculations with the PP1 and PP2 pseudopotentials
yield similar $\kappa_e$ values of around 480 and 770 ${\rm Wm^{-1}K^{-1}}$
at 0.5 and 1.0 eV, respectively.
The results are consistent with those from Knyazev {\it et al.},
Vlc\v{e}k {\it et al.} and Witte {\it et al.}
However, the $\kappa_e$ values from the two pseudopotentials at 5.0 eV deviate.
For instance, the result of DP-OF (PBE) with the PP2 pseudopotential is 1604.6 ${\rm Wm^{-1}K^{-1}}$
while the result with the PP1 pseudopotential is only 1281.7 ${\rm Wm^{-1}K^{-1}}$,
which is 20.1\% lower than the former one.
The $\kappa_e$ values from PP1 are close to those from Witte {\it et al.}, while
the $\kappa_e$ values from PP2 are consistent with the Knyazev {\it et al.} data.
Note that Witte {\it et al.} utilized a PAW potential with 11 valence electrons and the PBE XC functional,
while Knyazev {\it et al.} adopted an ultrasoft pseudopotential with 3 valence electrons
and the LDA XC functional.
It is also worth mentioning that a 64-atom cell is utilized in the work by Witte {\it et al.}
while a 256-atom cell is used in the work by Knyazev {\it et al.}
In this regard, the size effects may exist in the previous one according to our analysis.
In general, both values of electronic thermal conductivity lie within the experimental data of McKelvey {\it et al.}
Even so, our results demonstrate that different pseudopotentials may substantially affect
the results of $\kappa_e$ at high temperatures.
One possible reason that causes the deviation of $\kappa_e$ at 5.0 eV is the number of
electrons included in the pseudopotentials. However, it is also possible
that the deviation comes from the fact that
the non-local potential correction~\cite{91B-Read,07JPCM-Knider} in Eq.~(\ref{Ocoe}) is ignored.
Therefore, non-local corrections have to be considered when calculating the conductivity via the Kubo-Greenwood formula which is subject of future work.

\subsection{Ionic Thermal Conductivity}

\begin{figure}
	\centering
	\includegraphics[width=12cm]{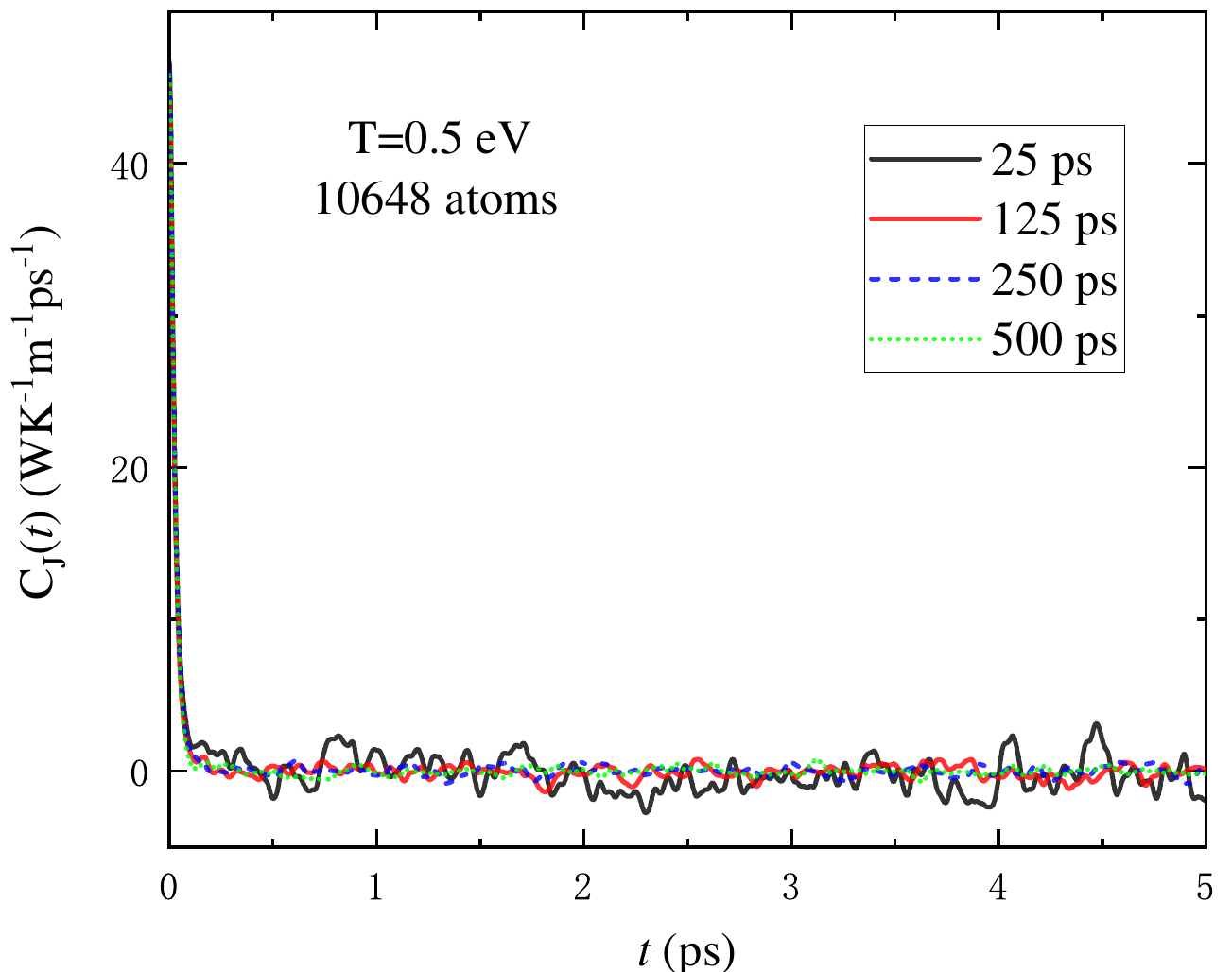}\\
	\caption{(Color online) Auto-correlation function of heat current $C_J(t)$ evaluated from
  different lengths of DPMD trajectories, i.e., 25, 125, 250 and 500 ps.
  The number of Al atoms in the cell is 10684 and the temperature is $T$=0.5 eV.
  The DPMD model was trained from OFDFT with the PBE exchange-correlation functional.
	}\label{fig:difft}
\end{figure}

The ionic thermal conductivity of warm dense Al can be evaluated by the GK formula
since the atomic energies are available in the DPMD method.
However, the computed ionic thermal conductivity may be affected by trajectory length and system size.
In this regard, we study the convergence of the ionic thermal conductivity
with respect to different lengths of trajectories and system sizes.
We first test the convergence of the auto-correlation function $C_J(t)$ in Eq.~(\ref{eq:Ct})
with respect to different lengths of trajectories and the results are shown in Fig.~\ref{fig:difft}.
A 10648-atom Al system was adopted with four different lengths of trajectories,
i.e., 25, 125, 250, and 500 ps, and the DPMD model was trained from OFDFT
with the PBE XC functional at a temperature of 0.5 eV.
As illustrated in Fig.~\ref{fig:difft},
the $C_J(t)$ curves abruptly decay within the first 0.1 ps
and oscillate with respect to time $t$.
We notice that the oscillations are largely affected
by the length of simulation time.
For example, $C_J(t)$ obtained from the 25-ps trajectory exhibits
substantially larger oscillations than the other three trajectories,
and the convergence is better achieved when the 250-ps trajectory is adopted.
The above results suggest that in order to yield converged $C_J(t)$ for warm dense Al,
a few hundreds of $ps$ are required even for a system with more than ten thousand atoms,
which is beyond the capability of FPMD simulations but can be realized by DPMD simulations.

\begin{figure}
	\centering
	\includegraphics[width=12cm]{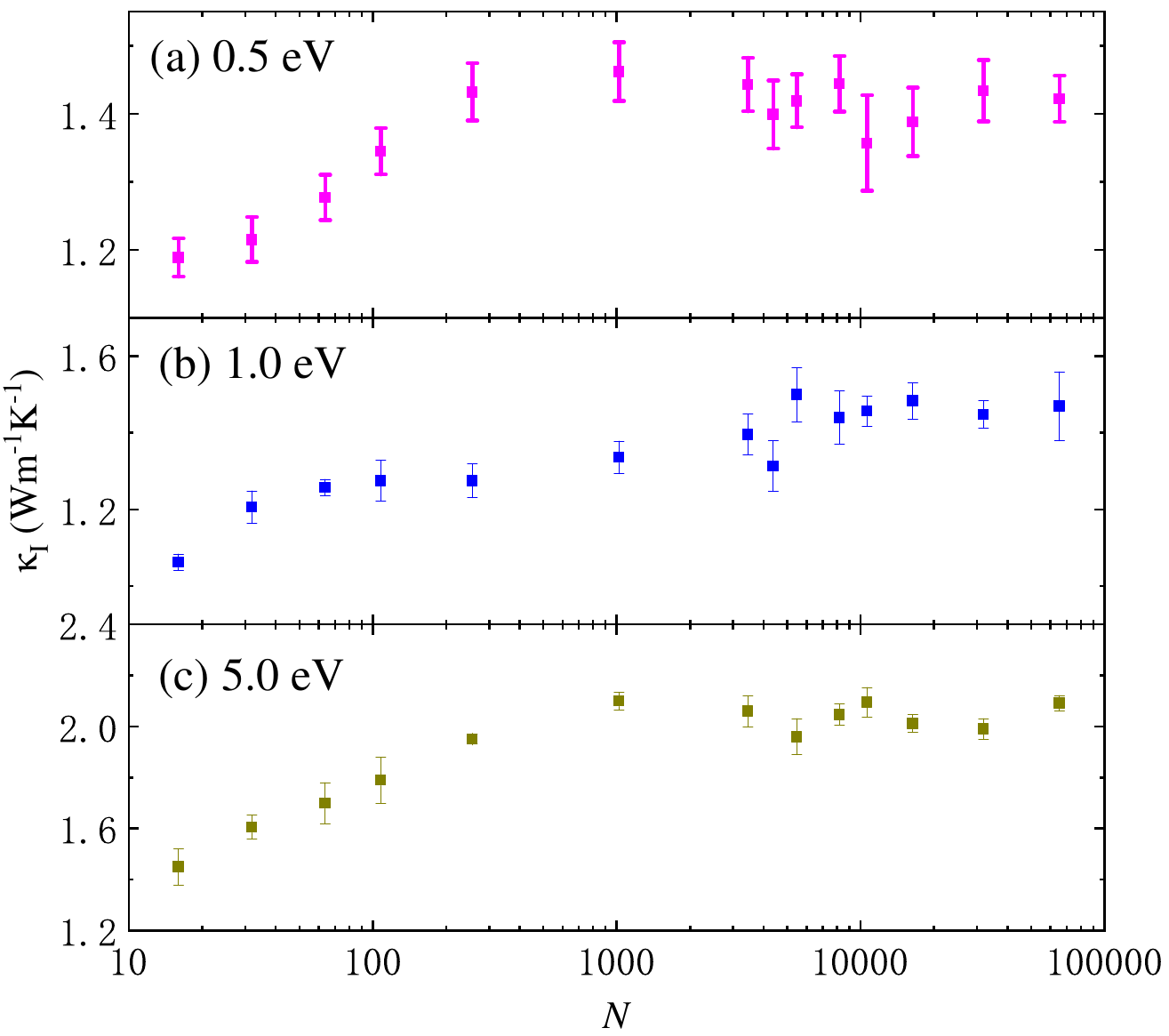}\\
	\caption{
(Color online) Computed ionic thermal conductivity of warm dense Al at (a) 0.5 eV, (b) 1.0 eV and (c) 5.0 eV with different sizes of systems. The number of atoms $N$ in different cells are
16, 32, 64, 108, 256, 1024, 5488, 8192, 10648, 16384, 32000 and 65536.
The results are obtained through DPMD trained from OFDFT with the PBE XC functional.}\label{fig:size_effect}
\end{figure}

Next, we run 12 different sizes of cells ranging from 16 to 65536 atoms for 500 ps
to check the size effects on ionic thermal conductivity,
and the results are illustrated in Fig.~\ref{fig:size_effect}.
Since $C_J(t)$ cannot strictly reach zero, a truncation of the correlation time $t$
is applied to the integration of $C_J(t)$ in Eq.~(\ref{kappa_i}).
In practice, we choose multiple truncations of $t$ ranging from 0.5 to 1.5 ps and
computed the error bars with the maximum and minimum integral values,
which are also shown in Fig.~\ref{fig:size_effect}.
We find that the ionic thermal conductivity increases with larger system sizes ranging
from 16- to 1024 atoms at all of the three temperatures considered for warm dense Al.
Next, the ionic thermal conductivity begins to oscillate until the
largest system size adopted (65536 atoms).
In this regard, we conclude that at least a 1024-atom system should be adopted and
a better converged ionic thermal conductivity can be obtained if a larger size of system
is used.
A 65536-atom cell is utilized to compute
the ionic thermal conductivity and the results are in Table \ref{tab:kappa}.
The ionic thermal conductivity of warm dense Al
is around 1-2 ${\rm Wm^{-1}K^{-1}}$,
which is more than two orders of magnitudes
smaller than the electronic thermal conductivity counterpart.
Additionally, both PBE and LDA XC functionals yield similar values
for the ionic thermal conductivity.

%Besides, the size of the simulated cell affects the ionic thermal conductivity
%much when $L$ is lower than 20 $ \rm \AA$ and the small cell underestimates the ionic thermal conductivity.

\section{Conclusions}
We propose a method that combines DPMD and DFT
to calculate both electronic and ionic thermal conductivities
of materials, and the DP models are trained from DFT-based MD trajectories.
The resulting DP models accurately reproduce the properties
as compared to those from DFT.
In addition, the DP models can be utilized to efficiently simulate
a large cell consisting of hundreds of atoms,
which largely mitigate the size effects caused by periodic boundary conditions.
Next, by using the atomic configurations from DPMD trajectories,
one can used the eigenvalues and eigenstates of a given system obtained from DFT solutions,
and adopt the Kubo-Greenwood formula to compute the electronic thermal conductivity.
In addition, the DP models yield atomic energies,
which are not available in the traditional DFT method.
By using the atomic energies to evaluate ionic thermal conductivity,
both electronic and ionic contributions to the thermal conductivity can be obtained for
a given material.

We took warm dense Al as an example and thoroughly
studied its thermal conductivity.
Expensive FPMD simulations of large systems
can be replaced by DPMD simulations with much smaller computational resources.
We first computed the temperature-dependent electronic thermal conductivities
of warm dense Al from 0.5 to 5.0 eV at a density of 2.7 ${\rm g/cm^3}$
with snapshots from OFDFT, KSDFT and DPMD,
and the three methods yielded almost the same results, demonstrating
that the DPMD method owns similar accuracy as FPMD simulations.
We then systematically investigated the convergence issues with respect to
the number of $k$-points, the number of atoms, the broadening parameter, the exchange-correlation functionals,
and the pseudopotentials. A 256-atom system was found to be large enough to converge the
electronic thermal conductivity. The broadening parameter was chosen to be 0.4 eV according
to our analysis of the energy interval distribution function. We found both
LDA and PBE XC functionals yielded similar results for the electronic thermal conductivity.
However, the choices of pseudopotentials may substantially affect the resulting electronic
thermal conductivity.
Furthermore, we also computed the ionic thermal conductivity with DPMD and the GK method,
and investigated the convergence issues with respect to trajectory length and system size.
We found the ionic thermal conductivity of warm dense Al is much smaller
than its electronic thermal conductivity.
In summary, the DPMD method provides a promising accuracy and efficiency in studying both
electronic and ionic thermal conductivity of warm dense Al
and should be considered for future work on modeling transport properties of WDM.

\newpage

\acknowledgements
This work was Supported by the Strategic Priority Research Program of Chinese Academy of Sciences Grant No. XDC01040100.
The work of M. C. is supported by the National Science Foundation of China under Grant No. 12074007.
The numerical simulations were performed on the High Performance Computing Platform of CAPT.

\bibliography{Al_WDM_cond}

\newpage

\newpage

\newpage

\newpage

\newpage

\newpage

\newpage

%\begin{figure}
%	\centering
%	\includegraphics[width=12cm]{fig/k-eidf.pdf}
%	\caption{(Color online) Distribution of energy intervals with different numbers of $k$-points adopted in KSDFT
%calculations. The cell contains 64 Al atoms and the temperature is 1.0 eV.
%	}\label{fig:k-eidf}
%\end{figure}

\clearpage
\newpage

\newpage

\newpage

\newpage

\clearpage
\newpage

\begin{table}[htbp]
	\centering
	\caption{ Sizes of $k$-points that adopted in KSDFT calculations to converge
the electronic thermal conductivity of Al with different number of atoms ($N$) in the simulation cell
at temperatures of 0.5, 1.0 and 5.0 eV.
	}
	\label{tab:K-points}
	\begin{tabular}{cccccc}
		\toprule
		\hline
		$N$ &0.5 eV&&1.0 eV&&5.0 eV\\
		\midrule
		\hline
		16 &$8\times8\times8$&&$7\times7\times7$&&$3\times3\times3$\\
		32 &$6\times6\times6$&&$5\times5\times5$&&$3\times3\times3$\\
		64 &$3\times3\times3$&&$3\times3\times3$&&$1\times1\times1$\\
		108&$2\times2\times2$&&$2\times2\times2$&&$1\times1\times1$\\
		216&$2\times2\times2$&&$2\times2\times2$&&$1\times1\times1$\\
		256&$2\times2\times2$&&$2\times2\times2$&&$1\times1\times1$\\
		432&$2\times2\times2$&& N/A       &&   N/A   \\
		\bottomrule
		\hline
	\end{tabular}
\end{table}

\clearpage
\newpage

\begin{table}[htbp]
	\centering
	\caption{Electronic thermal conductivity $\kappa_e$ (${\rm Wm^{-1}K^{-1}}$) and
ionic thermal conductivity $\kappa_I$ (${\rm Wm^{-1}K^{-1}}$) at temperatures $T$ of 0.5, 1.0 and 5.0 eV.
The results are computed from the DP-KS and DP-OF molecular dynamics trajectories.
DP-KS and DP-OF refer to the DP models trained from KSDFT and OFDFT molecular dynamics trajectories, respectively.
	}
	\label{tab:kappa}
\begin{tabular}{cccccccc}
	\toprule
	\hline
	&$T$ &\ \ &$\kappa_e$(PP1) &\ \ &$\kappa_e$(PP2) &\ \ &$\kappa_I$\\
	\midrule
	\hline
	            &0.5 eV& & 485.1& & 486.6&  &1.422$\pm0.034$\\
	DP-OF (PBE) &1.0 eV& & 764.1& & 772.3&  &1.469$\pm0.086$\\
	            &5.0 eV& &1281.7& &1604.6&  &2.091$\pm0.031$\\
   	            &0.5 eV& & 477.3& & 475.6&  &1.394$\pm0.047$\\
	DP-OF (LDA) &1.0 eV& & 773.8& & 779.0&  &1.318$\pm0.032$\\
	            &5.0 eV& &1246.4& &1568.5&  &2.141$\pm0.051$\\
 	            &0.5 eV& & 466.5& &   &  &1.419$\pm0.038$\\
	DP-KS (PBE) &1.0 eV& & 771.0& &   &  &1.393$\pm0.066$\\
	            &5.0 eV& &1305.8& &   &  &2.075$\pm0.066$\\
	\bottomrule
	\hline
\end{tabular}
\end{table}

\end{document}